\documentclass[conference]{IEEEtran}
\IEEEoverridecommandlockouts
\usepackage{cite}
\usepackage{amsmath,amssymb,amsfonts}
\usepackage{algorithmic}
\usepackage{graphicx}
\usepackage{textcomp}
\usepackage{xcolor}
\usepackage{subcaption}

\usepackage{upgreek}
\usepackage[colorlinks,
            linkcolor=blue,
            anchorcolor=blue,
            citecolor=green,
            ]{hyperref}

\def\BibTeX{{\rm B\kern-.05em{\sc i\kern-.025em b}\kern-.08em
    T\kern-.1667em\lower.7ex\hbox{E}\kern-.125emX}}
\begin{document}

\title{Lyapunov-guided Multi-Agent Reinforcement Learning for Delay-Sensitive Wireless Scheduling}

\author{
    \IEEEauthorblockN{Cheng Zhang\IEEEauthorrefmark{1}\IEEEauthorrefmark{2}, Lan Wei\IEEEauthorrefmark{1}\IEEEauthorrefmark{2}, Ji Fan\IEEEauthorrefmark{1}\IEEEauthorrefmark{2}, Zening Liu\IEEEauthorrefmark{2}, Yongming Huang\IEEEauthorrefmark{1}\IEEEauthorrefmark{2}}
    \IEEEauthorblockA{\IEEEauthorrefmark{1}National Mobile Communication Research Laboratory, Southeast University, Nanjing 210096, China\\
    \IEEEauthorrefmark{2}Purple Mountain Laboratories, Nanjing 211111, China\\
    Email: \{zhangcheng\_seu, weilan, 220240934, huangym\}@seu.edu.cn, liuzening@pmlabs.com.cn}
}

\maketitle

\begin{abstract}
In this paper, a two-stage intelligent scheduler is proposed to minimize the packet-level delay jitter while guaranteeing delay bound. Firstly, Lyapunov technology is employed to transform the delay-violation constraint into a sequential slot-level queue stability problem. Secondly, a hierarchical scheme is proposed to solve the resource allocation between multiple base stations and users, where the multi-agent reinforcement learning (MARL) gives the user priority and the number of scheduled packets, while the underlying scheduler allocates the resource. Our proposed scheme achieves lower delay jitter and delay violation rate than the Round-Robin Earliest Deadline First algorithm and MARL with delay violation penalty.
\end{abstract}


\section{Introduction}
With the advancement of 5G technology, the Ultra-Reliable Low Latency Communication (URLLC) scenario has garnered increasing attention. 
Within the URLLC scenario, there is a significant subset of applications related to industrial process control. 
According to 3GPP standards \cite{3gpp2021service}, in the closed-loop control application, the end-to-end (E2E) delay should be controlled within 2 ms, the reliability needs to achieve $1-10^{-6}$, and the delay jitter should be controlled to 1 $\upmu$s. 
Such applications impose stringent performance requirements on the communication systems, necessitating the assurance of reliability while minimizing the delay and jitter.

Related work has studied the resource allocation problem in delay-sensitive scenarios.
Yin et al. \cite{yin2023joint} proposed a long-term energy efficiency optimization scheme considering the queue conditions for the cell-free multiple-input multiple-output system in the URLLC scenario.
Addressing the joint scheduling and power allocation problem in a downlink cellular system, Ewaisha et al. \cite{ewaisha2017optimal} proposed two algorithms to cater to both real-time and non-real-time users. With the support of Lyapunov techniques, these algorithms achieve the system's capacity region.
Similarly, Neely et al. \cite{neely2013dynamic} developed a scheduling algorithm suitable for users with different delay constraints in a single-hop wireless communication system. 
Moreover, Reinforcement Learning (RL) has been widely applied to address the resource allocation problem in URLLC scenarios due to its strong adaptability to dynamic networks.
Shafieirad et al. \cite{shafieirad2022meeting} explored the challenge of scheduling under maximum delay constraints in downlink multi-cell wireless communication networks. The authors introduced a novel approach within an RL framework to tackle these constraints effectively.
Fan et al. \cite{fan2021delay} transformed the resource allocation problem into an integer nonlinear optimization problem, focusing on minimizing the delay of all tasks while ensuring the constraint of the Quality of Service (QoS) requirements, and an online RL strategy is further employed for the real-time dynamic resource allocation. 
Elsayed et al. \cite{elsayed2019reinforcement} investigated the coexistence of URLLC and enhanced Mobile Broadband users in 5G networks, and proposed a Q-learning-based RL algorithm optimizing the power and resource allocation.

In existing works, the impact of delay jitter is often neglected.
In the time-sensitive scenarios such as industrial automation and remote healthcare, the delay jitter can lead to unstable data transmission, affecting the system responsiveness and processing efficiency \cite{gutierrez2017synchronization}.
Furthermore, prior studies often treat delay violation as an optimization objective, aiming to minimize the probability of delay violation, rather than considering it as a strict constraint. Although Yin et al. \cite{yin2023joint} addressed delay violation as a constraint, their approach transformed the constraint on delay violation probability into a constraint on queue length. This transformation gradually relaxes the original constraint condition over an extended time window, introducing bias in practical applications.
Moreover, due to the substantial uncertainties inherent in practical communication systems, ensuring deterministic low delay has become exceedingly challenging. In aiming to ensure communication delay falls within a certain range with a specified probability, guaranteeing probabilistic delay holds significant value.

The contributions of this letter can be summarized as follows: 
Firstly, a multi-cell multi-user scenario requiring low delay jitter and ensuring probabilistic delay constraints is investigated. 
Secondly, a joint model and data-driven two-stage intelligent scheduling algorithm aimed at minimizing jitter with the probabilistic delay constraint is proposed. In the first stage, assisted by Lyapunov techniques, the long-term delay violation probability constraint is transformed into a single-slot virtual queue stability condition. In the second stage, to address the challenge of a large action space when directly allocating resources, a hierarchical algorithm is introduced. A multi-agent RL (MARL) algorithm is employed to allocate priorities and guide the number of packets to be scheduled for each user, based on which the underlying scheduler allocates resources to each user. 
Simulations show that compared to the Round-Robin Earliest Deadline First (EDF) algorithm and the MARL  with delay violation penalty, the proposed two-stage scheme perfectly guarantees bounded delay and effectively reduces delay jitter with medium-to-low traffic load. Under high traffic load, it also reduces delay jitter while achieves lower delay-bound violation.

\section{System Model and Problem Formulation}
In this section, we present the system model and problem formulation for the delay-sensitive communication scenario.

\subsection{System Model}

We consider a downlink Orthogonal Frequency Division Multiple Access (OFDMA) system consisting of $B$ base stations (BSs) and $U$ user equipment (UEs). Each BS is equipped with $M$ antennas, and each UE is equipped with a single antenna.
The frequency band is divided into $F$ orthogonal subcarriers. 
Denote the set of BSs as $\mathbb{B} = \{1, \cdots, B\}$. Similarly, $\mathbb{U} = \{1, \cdots, U\}$ represents the set of UEs, and $\mathbb{F} = \{1, \cdots, F\}$ denotes the set of subcarriers.
For each UE $u\in \mathbb{U}$, the index of its associated BS is denoted as $\bar{b}_u\in \mathbb{B}$. 
The set of UEs associated with BS $b$ is denoted as $\mathbb{U}_b$. For any $b, b' \in \mathbb{B}$, $\mathbb{U}_{b} \cap \mathbb{U}_{b'} = \varnothing $, where $b \neq b'$.
For the sake of simplicity, we assume that a resource block (RB) consists of a single subcarrier in the frequency domain and an OFDM symbol in the time domain. In addition, to focus on the scheduling problem, we assume that all BSs have perfect Channel State Information (CSI) of their associated UEs on all subcarriers. And the cooperation processing between multiple BSs is possible, e.g., under the control of central process unit, as in Cell-free systems.

Consider the downlink channel ${h}^m_{f,b,u}$ from the antenna $m$ of BS $b$ to UE $u$ on subcarrier $f$. It is affected by the large-scale fading factor $\beta_{b, u}$ and the small-scale fading factor $g^m_{f, b, u}$. ${h}^m_{f,b,u}$ can be calculated by ${h}^m_{f,b,u} = g^m_{f,b,u} \sqrt{\beta_{b, u}}$.
The distribution of $\beta_{b,u}$ and  $g^m_{f, b, u}$ can take various forms. For example, $g^m_{f,b,u}$ follows a zero-mean, unit variance complex Gaussian distribution, i.e., $g^m_{f,b,u} \sim \mathcal{C N}(0,1)$, and $\beta_{b, u}$ is related to the distance between BS $b$ and UE $u$, $d_{b, u}$, i.e. $\beta_{b, u} = \left(1+{d_{b, u}}/{d_0}\right)^{-{\alpha}}$, where $d_0$ is the reference distance of the path loss model, ${\alpha}$ is the path loss exponent. Thus, the channel gain from BS $b$ to UE $u$ on subcarrier $f$ can be represented as $\boldsymbol{h}_{f,b,u} = [{h}^1_{f,b,u}, \cdots, {h}^M_{f,b,u}]^{\text{T}}$.

Considering each BS can schedule one user on each available subcarrier in each time slot, binary variables $\zeta_{u, f}^t$ are used to indicate whether user $u$ is scheduled on subcarrier $f$ in time slot $t$. In other words, in time slot $t$, if subcarrier $f$ is allocated to user $u$, then $\zeta_{u, f}^t = 1$, otherwise $\zeta_{u, f}^t = 0$.
The beamforming vector of BS $\bar{b}_u$ for its associated UE $u$ on subcarrier $f$ in time slot $t$ is defined as $\boldsymbol{w}^t_{f,\bar{b}_u,u}$.
Based on the above model, the Signal-to-Interference-plus-Noise Ratio (SINR) $\gamma_{u, f}^t$ for UE $u$ on subcarrier $f$ in time slot $t$ can be calculated as Eq. \eqref{eq: sinr_gamma},
\begin{align}\label{eq: sinr_gamma}
\gamma_{u, f}^t = \frac{\zeta_{u, f}^t ||{\boldsymbol{h}^{t, \text{H}}_{f,\bar{b}_u,u}} \boldsymbol{w}^t_{f,\bar{b}_u,u} ||^2}
{{\sum_{v \in \mathbb{U}-{u}}} \zeta^{t}_{v,f} ||{\boldsymbol{h}^{t, \text{H}}_{f,\bar{b}_v,u}} \boldsymbol{w}^t_{f,\bar{b}_v,v} ||^2 + {\sigma^{t}_{u,f}}^2}, 
\end{align}
where $\sigma^{t}_{u,f}$ is the noise power for UE $u$ on subcarrier $f$ in time slot $t$.
Considering practical finite block-length coding \cite{polyanskiy2010channel}, the achievable rate of UE $u$ in time slot $t$, given a block error rate $\epsilon_{u}$, is:
\begin{equation}
    \psi_{u}^{t}=\sum_{f \in\mathbb{F}} \log_{2} \left(1 + \gamma_{u, f}^t \right) - Q^{-1} \left(\epsilon_{u}\right) \sqrt{ \sum_{f \in\mathbb{F}} V_{u, f}^{t}},
\end{equation}
where $Q^{-1}\left( \cdot \right)$ is the inverse function of the Gaussian Q function, and $V_{u, f}^{t} = \left[\log_{2} \left(e\right)\right]^{2} \left( 1 - \left(1+ \gamma_{u, f}^t \right) ^{-2}\right)$ is the channel dispersion.

\subsection{Network Queueing and Delay}

Denote the number of packets that arrive in time slot $t$ as $A_u(t)$, and we consider that the packets arrive in time slot $t$ can only be scheduled in time slots $t+1$ and thereafter \cite{mao2017stochastic}.
Denote $\mathbb{A}_u(t) = \{1, \cdots, A_u(t)\}$.
The size of each packet for user $u$ is represented by $G_{u}$ (bits).
The general assumption is that $A_{u}(t)$ follows a Poisson distribution with an expectation of $\lambda_u$. 
For the same user $u$, all arriving packets have the same deadline, denoted as $D_{u}$. 
We assume that the delay requirements of all users are known at the BS.
Each packet from UE $u$ is delay-sensitive and must be transmitted before its deadline $D_u$, 
and if it is not transmitted by the deadline, it is considered to violate the delay constraint. 
Taking the EDF principle into consideration, packets in each user's buffer should be scheduled in a First-In-First-Out (FIFO) manner. 
Denote the queue length as $Z_{u}(t)$ for user $u$ at the beginning of time slot $t$. Then, we have:
\begin{align}\label{eq: queue_length}
    Z_{u}(t + 1) = \left[ Z_{u}(t)-\psi_{u}^{t} \right]^{+} + A_{u}(t)G_{u},
\end{align}
where $\left[ x \right]^{+} \triangleq \max \{ x,0 \}$. 
For the packet with index $\chi_{u,t}^a$ which is the $a$-th packet arriving in time slot $t$ for user $u$, the delay can be denoted as:
\begin{align}\label{eq: delay}
  d_{\chi_{u,t}^a} = & \underset{\tau}{\arg}\min \Big \{ \left[Z_{u}(t)-\psi_{u}^{t}\right]^{+}+G_{u}a \nonumber\\
  & \hspace{1.4cm}- \sum_{i=t+1}^{i=\tau}\psi_{u}^{i}\le 0\Big \} - t. 
\end{align}

\subsection{Problem Formulation}
Based on the given system model, our objective is to design a joint resource allocation and scheduling algorithm to minimize the user delay jitter within a time window $T$, while ensuring that each user's packets are transmitted before their deadlines.

In the following, $\big \{ \zeta_{u, f}^t, u \in \mathbb{U}, f \in \mathbb{F}$, $t \in {\mathbb{T}} = \{1, \cdots, T\} \big \}$ is briefly denoted as $\left\{ \zeta_{u,f}^{t} \right\}$.
The optimization problem $\mathcal{O}1$ can be formulated as follows:
\begin{align} 
        & \mathcal{O}1: \min_{\{ \zeta_{u,f}^{t}\}} f\left(\left\{  \zeta_{u,f}^{t}\right\}\right) \label{eqo1: target} \\
        \text{s.t.} & \quad \text{Pr}\left(d_{\chi_{u,t}^a}>D_{u}\right)<\eta_{u},u\in\mathbb{U},a \in \mathbb{A}_u(t),t\in{\mathbb{T}} \label{eqo1: delay_violate} \\
        & \quad  \zeta_{u,f}^{t} \in \left\{0, 1\right\} , u \in \mathbb{U}, t \in \mathbb{T}, f\in\mathbb{F} \label{eqo1: zeta} \\
        & \quad \sum_{u \in \mathbb{U}_b}\zeta_{u,f}^{t} \le 1, t \in \mathbb{T}, b\in \mathbb{B}, f\in\mathbb{F} \label{eqo1: one_subcarrier} \\
        & \quad \sum_{u \in \mathbb{U}_b} \sum_{f \in \mathbb{F}} {\Vert \boldsymbol{w}_{f,b,u}^{t} \Vert}^2 \zeta_{u, f}^t \leq P_{b, \text{max}}, t \in \mathbb{T}, b \in \mathbb{B}, \label{eqo1: power_constraint}
\end{align}
where $\eta_{u}$ is the upper bound on the delay violation probability for user $u$, $P_{b, \text{max}}$ denotes the maximum power of BS $b$. The system average delay jitter is expressed as in Eq. \eqref{eq: jitter}:
\begin{align}\label{eq: jitter}
f\left(\left\{ \zeta_{u, f}^t\right\}\right) = \frac{1}{U} \sum_{u \in \mathbb{U}} \sqrt{ \frac{1}{\vert \mathbb{C}_u \vert} \sum_{c \in \mathbb{C}_u} \left( d_{c} - \frac{1}{\left| \mathbb{C}_u \right|} \sum_{c \in \mathbb{C}_u} d_{c} \right)^2},
\end{align}
where $\mathbb{C}_u$ denotes the index set of scheduled packets for user $u$ during the time window $\mathbb{T}$.

\section{Lyapunov-guided QMIX-powered Intelligent Packet Scheduling}
Considering that constraint (\ref{eqo1: delay_violate}) is a probabilistic constraint and the distribution of random variables is unknown, directly solving problem $\mathcal{O}1$ is very challenging. Additionally, problem $\mathcal{O}1$ is a multi-slot stochastic optimization problem with constraints where few effective solutions are available. In the following, we propose a two-stage intelligent scheduling algorithm, namely the Lyapunov-guided QMIX-powered Intelligent Packet Scheduling (LGQP-IPS) algorithm. 

In the first stage, a Lyapunov optimization approach is employed to transform the probabilistic delay constraint into a virtual queue stability condition. 
By minimizing the Lyapunov drift, which characterizes the changes in system backlog, the actual and virtual queues can be stabled, thereby ensuring delay constraint. And minimizing the penalty term, i.e., delay jitter, the optimization objective can be achieved. 
Considering that the complexity of this sequential single-slot resource allocation problem grows exponentially with the number of users, we further introduce a hierarchical QMIX-based multi-agent RL algorithm to address this issue, where $U$ agents are constructed to make the decision for each user. Specifically, we transform the resource allocation problem into a scheduling problem. 
In the learning layer, the intelligent algorithm provides priorities for each user and the number of data packets to be transmitted. 
In the scheduling layer, the scheduler assigns subcarriers to each user according to their priority until all resources are allocated or depleted.

\begin{figure}[htb]
  \centerline{\includegraphics[scale=0.5]{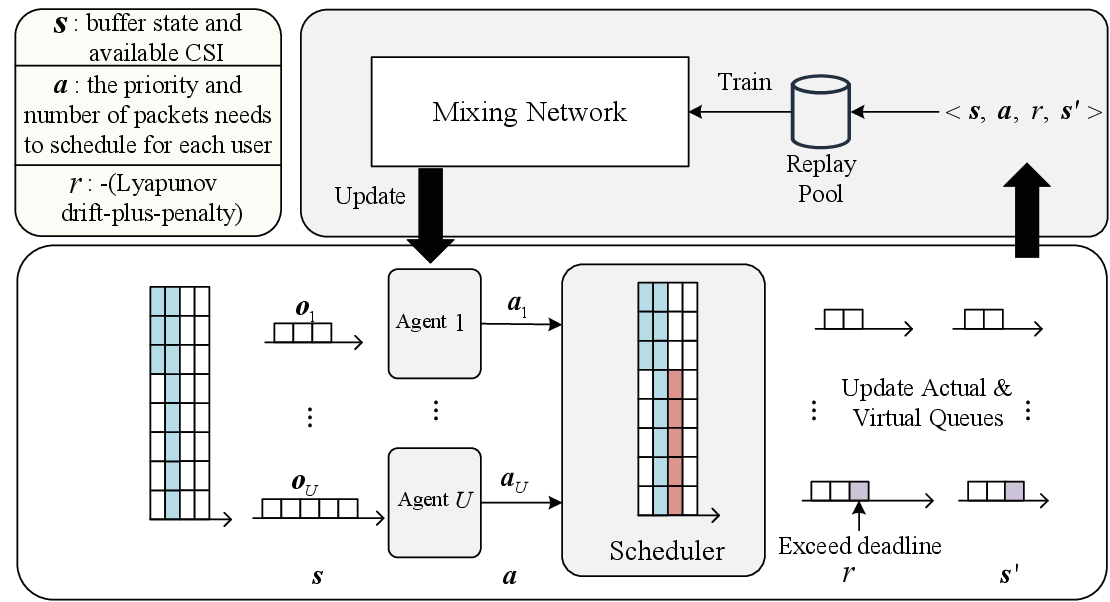}}
  \caption{The process of the proposed intelligent packet scheduling algorithm.}
  \label{fig: system_fructure}
\end{figure}

The overall process of this algorithm is illustrated in Fig. {\ref{fig: system_fructure}}. 
Specifically, firstly, guided by the Lyapunov technique, a virtual queue model is established, and the expression of the upper bound of Lyapunov drift-plus-penalty is calculated for each slot $t$. 
Secondly, at each slot $t$, given the  observation $\boldsymbol{o}_{u}(t)$ consisting of the buffer state of user $u$ 
at the beginning of the time slot $t$, i.e., $Z_u(t)$, and its current available CSI, the action, i.e., $\boldsymbol{a}_u(t)$,
with the maximum Q-value is determined for each user $u$ by each agent $u$, thus obtaining the priority of the user and the number of packets to be scheduled. 
Subsequently, the scheduler allocates subcarriers to each user in order of priority until the data packets to be transmitted can be accommodated. 
Finally, the reward $r$ is calculated guided by drift-plus-penalty, and each user's buffer status is updated while packet timeout situations are recorded. 
The Mixing Network of QMIX collects the quadruple $<\boldsymbol{s}, \boldsymbol{a}, r, \boldsymbol{s}'>$, where $\boldsymbol{a}=[\boldsymbol{a}_1,\boldsymbol{a}_2,...,\boldsymbol{a}_{U}]$ and $\boldsymbol{s}=[\boldsymbol{o}_1,\boldsymbol{o}_2,...,\boldsymbol{o}_{U}]$
denote the joint action and observation, and places it into the Replay Pool for training and updating the agent's network.

In Section \ref{sec: delay_constraint}, we will provide a detailed explanation of the transformation of delay constraint and an in-depth explanation of the QMIX-powered scheduling design will be given in Section \ref{sec: marl}.

\subsection{Delay Constraint Transformation}
\label{sec: delay_constraint}

Denote the survival time of packet $\chi_{u,t}^a$ until time slot $\tau$ as $p_{\chi_{u,t}^a}(\tau)$. Let $\omega_{\chi_{u,t}^a}(\tau)$ be the delay violation indicator at the beginning of time slot $\tau$, i.e., $\omega_{\chi_{u,t}^a}(\tau) = 1$ if $p_{\chi_{u,t}^a}(\tau) = D_u + 1$, otherwise $\omega_{\chi_{u,t}^a}(\tau) = 0$.
To avoid double counting, we consider $\omega_{\chi_{u,t}^a}(\tau') = 0$ for slot $\tau'$ for $\tau' > \tau$ if $\omega_{\chi_{u,t}^a}(\tau) = 1$.

According to the law of large numbers, i.e., as the number of experiments increases, the sample mean approaches the population mean more closely, we consider an infinite observation time to re-express the delay violation constraint in Eq. \eqref{eqo1: delay_violate} as
\begin{equation} \label{eq: big_num_1}
  \frac{\lim_{T \rightarrow\infty} \sum_{t =1}^T \sum_{\tau = 1}^{t}\sum_{a \in \mathbb{A}_u(\tau)} \omega_{\chi_{u,\tau}^a}(t)}{\lim_{T\rightarrow \infty}\sum_{t=1}^{T}A_u(t)} < \eta_u,
\end{equation}
which can be rewritten as Eq. (\ref{eq: big_num_2}): 
\begin{equation}\label{eq: big_num_2}
    \lim_{{T \to \infty}} \sum_{{t = 1}}^T \left(\sum_{\tau = 1}^{t}\sum_{a \in \mathbb{A}_u(\tau)} \omega_{\chi_{u,\tau}^a}(t) - \eta_u A_u(t) \right) < 0.
\end{equation}
Eq. (\ref{eq: big_num_2}) can be a necessary condition for queue stability.
Therefore, a virtual queue $H_u(t)$ can be constructed, with the update criterion given by Eq. (\ref{eq: virtual_queue}):
\begin{equation} \label{eq: virtual_queue}
	H_u(t+1) = [H_u(t) - \eta_u A_u(t)]^{+} + \sum_{\tau = 1}^{t}\sum_{a \in \mathbb{A}_u(\tau)} \omega_{\chi_{u,\tau}^a}(t).
\end{equation}
Next, using Lyapunov techniques, the Lyapunov drift is first computed, which represents the change in the backlog of queues in the system.

Define $\boldsymbol{Z}(t) = [Z_1(t), \cdots, Z_U(t)]$, $\boldsymbol{H}(t) = [H_1(t), \cdots, H_U(t)]$, $\boldsymbol{V}(t) = [\boldsymbol{Z}(t) \ \boldsymbol{H}(t)]$, then the backlog of the system can be represented as:
  $L(\boldsymbol{V}(t)) \triangleq \frac{1}{2}\sum_{u=1}^{U}Z^2_u(t)+\frac{1}{2}\sum_{u=1}^{U}H_u^2(t)$.
The Lyapunov drift $\Delta_t$ in time slot $t$ can then be computed as follows:
\begin{align} \label{eq: drift}
  \begin{split}
      \Delta_t & = \mathbb{E}\left \{L(\boldsymbol{V}(t+1)) - L(\boldsymbol{V}(t)) | \boldsymbol{V}(t)\right\} \\
      & \leq B + \sum_{u=1}^U Z_u(t)\mathbb{E}\left\{G_uA_u(t)-\psi_u^t|\boldsymbol{V}(t)\right\} \\
      & \quad + \sum_{u=1}^U H_u(t) \mathbb{E}\left\{\sum_{\tau = 1}^{t}\sum_{a \in \mathbb{A}_u(\tau)} \omega_{\chi_{u,\tau}^a}(t) - \eta_u A_u(t)|\boldsymbol{V}(t) \right\} \\
      & = B + \sum_{u=1}^U Z_u(t)\left(G_u \lambda_u - \psi_u^t\right)\\
      & \quad + \sum_{u=1}^U H_u(t) \left(\sum_{\tau = 1}^{t}\sum_{a \in \mathbb{A}_u(\tau)} \omega_{\chi_{u,\tau}^a}(t) - \eta_u \lambda_u \right),
    \end{split}
\end{align}
where $B$ is a constant term. 
Based on the Lyapunov drift, the Lyapunov optimization problem $\mathcal{O}2$ for each time slot $t$ can be formulated as follows:
\begin{align}
      &\mathcal{O}2: \min_{\zeta_{u, f}^t, u\in\mathbb{U}, f\in\mathbb{F}} \Delta_t + \mu \bar{f}\left(\zeta_{u, f}^t, u\in\mathbb{U}, f\in\mathbb{F}\right) \label{eqo2: reward}\\
      \text{s.t.}
      & \quad \zeta_{u,f}^{t} \in \left\{0,1\right\} ,u \in \mathbb{U}, f\in\mathbb{F} \label{eqo2: zeta}\\
      & \quad \sum_{u \in \mathbb{U}_b}\zeta_{u, f}^{t} \le 1, b\in\mathbb{B}, f\in\mathbb{F} \label{eqo2: one_subcarrier} \\
      & \quad \sum_{u \in \mathbb{U}_b} \sum_{f \in \mathbb{F}} {\Vert \boldsymbol{w}_{f,b,u}^{t} \Vert}^2 \zeta_{u, f}^t \leq P_{b, \text{max}}, b \in \mathbb{B},
\end{align}
where $\mu$ is the penalty coefficient, which is used to balance the trade-off between delay jitter and delay violation probability. $\bar{f}(\cdot)$ is the current average delay jitter, which can be calculated from Eq. \eqref{eq: jitter} via replacing $\mathbb{C}_u$ with the index set of scheduled packets for user $u$ in time slots $1$ to $t$.

\subsection{QMIX for Intelligent Packet Scheduling}
\label{sec: marl}
QMIX \cite{rashid2020monotonic, zhu2022survey} stands out as a typical algorithm for multi-agent coordination. Its centralized training with a decentralized execution paradigm \cite{langer2020distributed} is particularly suited for scenarios that demand autonomous agent operation coupled with global coordination to optimize resource allocation dynamically. 
Observing Eq. (\ref{eqo2: zeta}), it is evident that RB allocation in different cells is interdependent, and cooperation among cells is necessary to achieve global optimality. Failure to effectively model the cooperative relationships between cells may lead to a local optimum.
Based on the aforementioned considerations, in the second stage of the LGQP-IPS algorithm, we adopted a hierarchical structure. In the learning layer, the QMIX algorithm provides priorities for each user and the number of data packets to be scheduled. In the scheduling layer, the scheduler allocates resource blocks (RBs) to users based on the output from the learning layer, prioritizing users until all resources are allocated or all users are assigned.

In our problem, we model the Markov decision process (MDP) problem as follows:
\begin{itemize}
  \item State $\boldsymbol{s}=[\boldsymbol{o}_1,\boldsymbol{o}_2,...,\boldsymbol{o}_{U}]$: The backlog of all users' buffers (in bits) and their available CSI.
  \item Action $\boldsymbol{a}=[\boldsymbol{a}_1,\boldsymbol{a}_2,...,\boldsymbol{a}_{U}]$: The priority of users and the number of packets to be transmitted.
  \item Reward $r$: A rewritten form based on the negative of the Lyapunov drift-plus-penalty term at time slot $t$.
\end{itemize}

In order to maintain the stability throughout the training process, the reward should be maintained in an appropriate range to prevent the Q-network from overestimating the Q-values, which requires suitable scaling and bias adjustment. The Lyapunov drift defined in Eq. (\ref{eq: drift}) is the combination of two terms separately calculated according to the actual queue $\boldsymbol{Z}(t)$, which counts by bits, and the virtual queue $\boldsymbol{H}(t)$, which counts by the number of packets. It is necessary to normalize the actual queue term in the Lyapunov drift by packet size:
\begin{align} \label{eq: new_drift}
  \begin{split}
      \Delta_t^{new} & = B + \sum_{u=1}^U \frac{Z_u(t)}{G_u}\left( \lambda_u - \frac{\psi_u^t}{G_u}\right)\\
      & \quad + \sum_{u=1}^U H_u(t) \left(\sum_{\tau = 1}^{t}\sum_{a \in \mathbb{A}_u(\tau)} \omega_{\chi_{u,\tau}^a}(t) - \eta_u \lambda_u \right),
    \end{split}
\end{align}
Thus, the reward in this work is designed as follows:
\begin{align} \label{eq: reward}
  \begin{split}
      r = - \left(\frac{\Delta_t^{new}(\boldsymbol{s}, \boldsymbol{a}) + \mu \bar{f}(\boldsymbol{s}, \boldsymbol{a})}{\Omega} - bias\right),
    \end{split}
\end{align}
where $\Omega$ is the scaling factor, and $\epsilon$ is the bias constant.

For each agent $u$, there's an individual value function 
$Q_u(\boldsymbol{\tau}_u(t), \boldsymbol{a}_u(t); \boldsymbol{\theta}_u)$, where $\boldsymbol{\tau}_u(t)$ is the action-observation history of agent $u$, i.e. $[(\boldsymbol{a}_u(0), \boldsymbol{o}_u(0)), \cdots, (\boldsymbol{a}_u(t-1), \boldsymbol{o}_u(t-1))]$, $\boldsymbol{a}_u(t)$ is the action taken by the agent $u$ in state $\boldsymbol{s}(t)$, and $\boldsymbol{\theta}_u$ are the parameters of the agent network.
The global Q-value is calculated by the mixing network, expressed as:
\begin{align} \label{eq: QMIX_total_Q_value}
  Q_\text{tot}(\boldsymbol{\tau}, \boldsymbol{a}; \boldsymbol{\theta}) = & \nonumber \\ 
  & \hspace{-1.5cm} g(Q_1(\boldsymbol{\tau}_1, \boldsymbol{a}_1; \boldsymbol{\theta}_1), \cdots, Q_U(\boldsymbol{\tau}_U, \boldsymbol{a}_U; \boldsymbol{\theta}_U); \boldsymbol{\phi}(\boldsymbol{s})),
\end{align}
where $g$ is the mixing network, which is a monotonic non-decreasing function of the individual agent's Q-values, ensuring consistency of the global $argmax$ operation. $\boldsymbol{\phi}$ is a hyper-network parameterized by the environmental state $\boldsymbol{s}$, and $\boldsymbol{\theta}$ is the set of parameters including all agents and the mixing network.
The weights in the mixing network are produced by the hyper-network, which controls the weights in the mixing network through the state $\boldsymbol{s}$, and all weights are non-negative to satisfy the monotonicity constraint.

QMIX uses the Squared Temporal Difference (TD) Error error as the loss function to train the parameters of the mixing network and the agent networks, which is shown in Eq. (\ref{eq: QMIX_loss}).
\begin{equation} \label{eq: QMIX_loss}
  \mathcal{L}(\boldsymbol{\theta}) = \mathbb{E}_{(\boldsymbol{s}, \boldsymbol{a}, r, \boldsymbol{s}') \sim \mathcal{D}} \left[ \left( y - Q_\text{tot}(\boldsymbol{\tau}, \boldsymbol{a}; \boldsymbol{\theta}) \right)^2 \right],
\end{equation}
where $y = r + \gamma \max_{\boldsymbol{a}'} Q_\text{tot}(\boldsymbol{\tau}', \boldsymbol{a}'; \boldsymbol{\theta}^{-})$, $\boldsymbol{\theta}^{-}$ are the parameters of the target network, $\boldsymbol{\tau}'$ and $\boldsymbol{a}'$ denote the action-observation history and joint action of next state, respectively, $\mathcal{D}$ is replay pool, and $\gamma$ is the discount factor.

Finally, parameters are updated using gradient descent:
$\boldsymbol{\theta} \leftarrow \boldsymbol{\theta} + \beta \nabla_{\boldsymbol{\theta}} \mathcal{L}(\boldsymbol{\theta})$,
where $\beta$ is the learning rate, and $\nabla_\theta \mathcal{L}(\theta)$ is the gradient of the loss function with respect to the parameters.

\section{Numerical Results}
In this section, numerical results that evaluate the performance of the proposed Lyapunov-guided MARL algorithm for delay-sensitive tasks will be presented.


\begin{figure}
  \centering
    \includegraphics[width=0.45\textwidth]{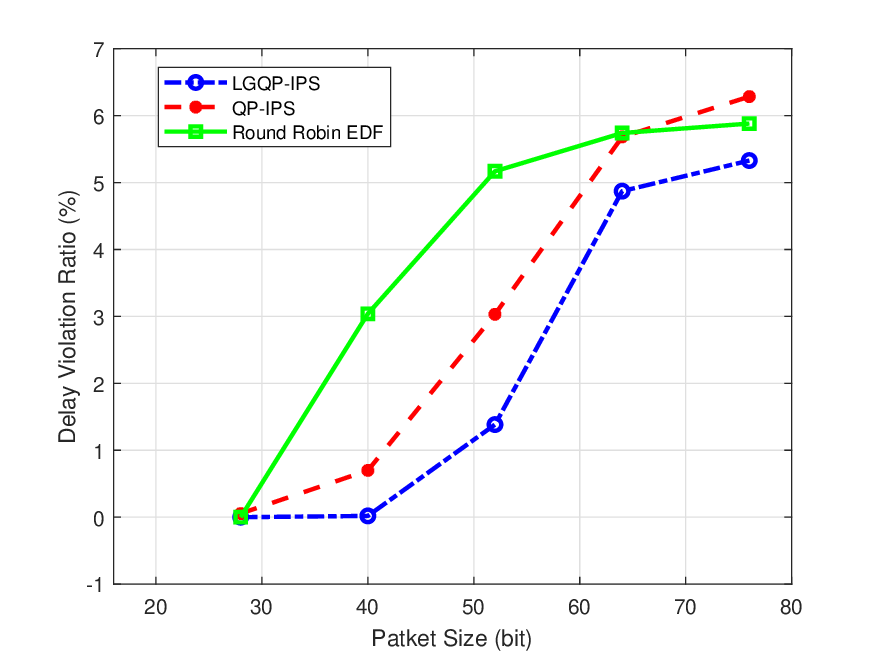} 
    \caption{The comparison of delay violation ratio between LGPS-IPS and baselines;} 
  \label{fig:performance_evaluation-1}
\end{figure}

\begin{figure}
  \centering
    \includegraphics[width=0.45\textwidth]{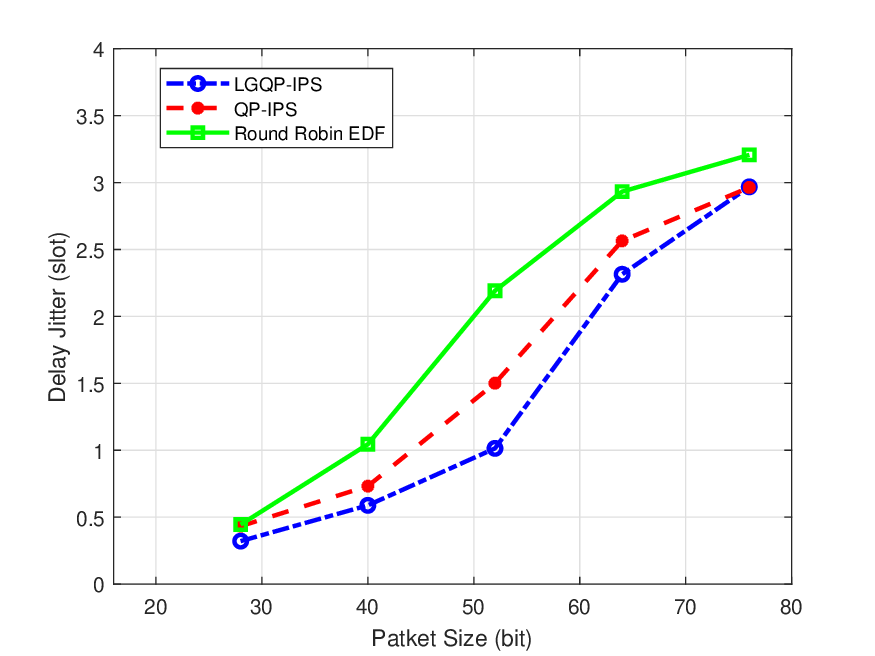}
      \caption{The comparison of jitter between LGPS-IPS and baselines;}
  \label{fig:performance_evaluation-2}
\end{figure}

\begin{figure*}[ht]
    \centering
    \begin{subfigure}{0.32\textwidth}
        \includegraphics[width=\linewidth]{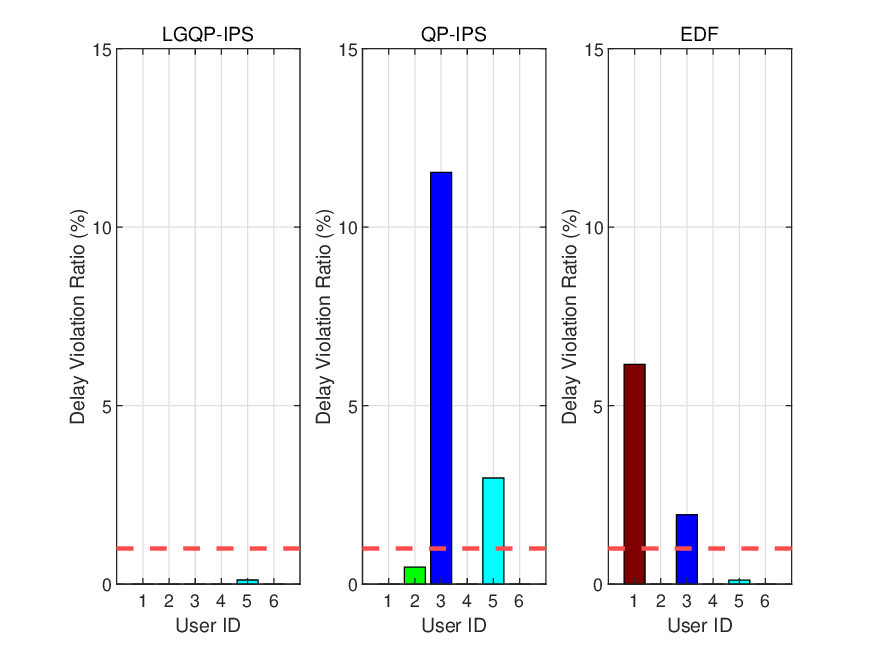} 
        \caption{$G_u$ = 40 bits}
        \label{fig:sub1}
    \end{subfigure}
    \hfill
    \begin{subfigure}{0.32\textwidth}
        \includegraphics[width=\linewidth]{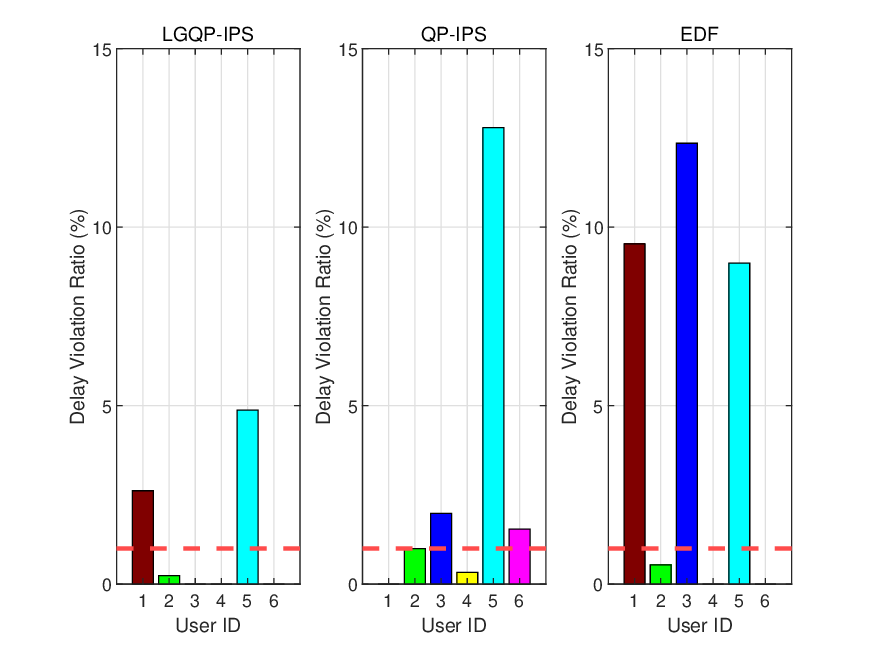}
        \caption{$G_u$ = 52 bits}
        \label{fig:sub2}
    \end{subfigure}
    \hfill
    \begin{subfigure}{0.32\textwidth}
        \includegraphics[width=\linewidth]{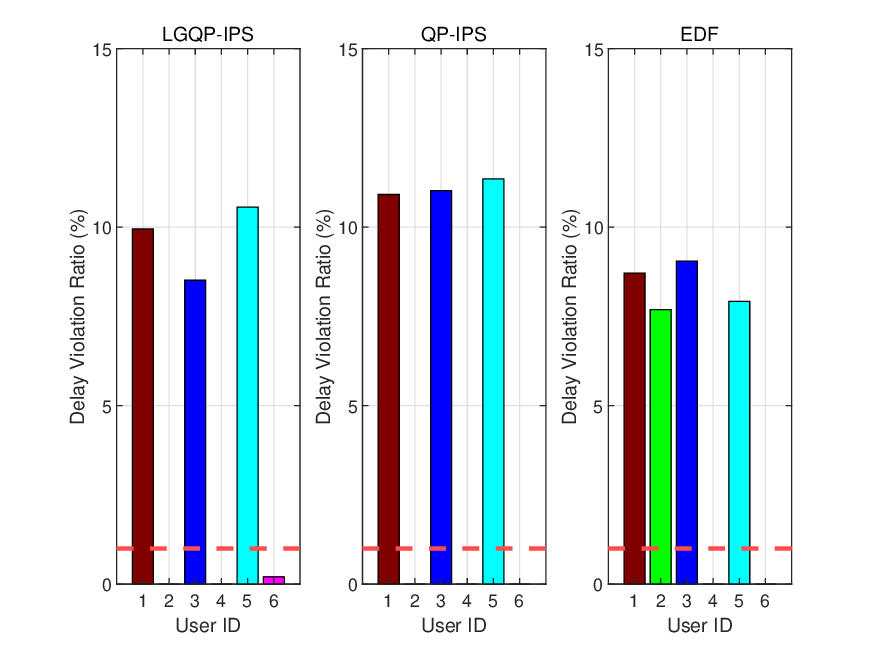}
        \caption{$G_u$ = 64 bits}
        \label{fig:sub3}
    \end{subfigure}
    \caption{The delay violation ratio of different users with different $G_u$}
    \label{fig:overall}
\end{figure*}

\subsection{Simulation Setup}
Consider a downlink OFDMA system with $3$ cells, $2$ UEs per cell, and $32$ subcarriers. 
Maximum Ratio Transmission (MRT) precoding is employed at the BSs, and the number of antennas at each BS is set to $16$.
The path loss exponent is set to $\alpha = 2$, the radius of the cell is $40$ m, and the reference distance is $d_0 = 1$ m. The noise power per subcarrier with the subcarrier bandwidth of 30 kHz is set to $-129$ dBm. 
$P_{b, \max}$ is set to $10$ w.
The packet size $G_u$ is set as $28$, $40$, $52$, $64$, $76$ bits, representing different traffic loads. The deadline for each packet is $5$, $2$, $5$, $3$, $2$, and $2$ time slots perspectively for each user. The observation window length $T = 50$ slot. The arrival rate $\lambda_u$ is set to $3$ for all of the users.
The block error rate $\epsilon_{j_b}$ is set to $10^{-9}$. 
The penalty coefficient $\mu$ as referred in Eq. (\ref{eqo2: reward}) is set to $50$, and the delay violation probability $\eta_u$ is set to $1$\%. The scaling factor $\Omega$ is set to $500$, and $bias$ is set to $1$. The learning rate $\beta$ is set to $0.0005$, and the discount factor $\gamma$ is set to $0.85$. 
The batch size and buffer size are set to $4096$ and $50000$.
The number of episodes is set to $2000$, and the number of time slots in each episode is set to $50$. The neural network for QMIX has $3$ layers with $64$ neurons per layer. The simulation is implemented in Python using the PyMARL library \cite{samvelyan19smac}.

\subsection{Performance Evaluation}
To evaluate the performance of the proposed algorithm, we compare it with the following baseline algorithm.
\begin{itemize}
  \item \textbf{QP-IPS}: A QMIX-powered intelligent packet scheduling (QP-IPS) algorithm with a penalty term for delay violation probability.
  \item \textbf{Round-Robin EDF}: A Round-Robin-based packet scheduling algorithm with EDF principle.
\end{itemize}

Specifically, in QP-IPS, the reward is set as 
  $r'(t) = - \left(\bar{f}(\boldsymbol{s}, \boldsymbol{a}) + \delta \frac{\sum_{m=1}^{t}\sum_{\tau = 1}^{m} \sum_{a \in \mathbb{A}_u(\tau)}  \omega_{\chi_{u,\tau}^a}(m)}{\sum_{m=1}^{t} A_u(m)}\right)$, 
where $ \delta $ represents the penalty coefficient, which is used to balance optimization objectives and delay violation rates. Actions and states remain consistent with LGQP-IPS. 
While Round-Robin EDF refers to selecting scheduled users in a round-robin manner, with each user transmitting its most urgent packet according to the EDF principle. If there are remaining resources, the scheduling proceeds to the next round.


The performance metrics used for evaluation are the average delay jitter and the delay violation probability. The results are shown in Fig. \ref{fig:performance_evaluation-1} and Fig. \ref{fig:performance_evaluation-2}.

Fig. \ref{fig:performance_evaluation-1} shows the delay violation ratio percentage over different packet size which represents different traffic loads. With relatively low traffic load, i.e., $G_u = 28$ bits, all three algorithms can carry the traffic load without any delay violation. As the traffic load increases, i.e., $G_u = 52$ bits, both LGQP-IPS and QP-IPS algorithms demonstrate a substantially lower violation ratio compared to the Round Robin EDF algorithm, and the LGQP-IPS algorithm exhibits a comparatively moderate rise, suggesting a better adaptability to high traffic loads. For higher loads, i.e., $G_u = 76$ bits, all three algorithms show poor performance, because the load is over the upper bound of system capacity.

Fig. \ref{fig:performance_evaluation-2} shows the variations in jitter, measured in slots, also across different packet size. 
Initially, under low load conditions, both LGQP-IPS and QP-IPS algorithms manage to reduce jitter to lower levels, i.e., when $G_u = 40$ bits both algorithms keep mean jitter under 0.8 slot, with LGQP-IPS achieving the most significant reduction. 
Under high load conditions, while the jitter for LGQP-IPS and the baseline algorithms increase, LGQP-IPS shows a rather lower jitter.
Again, if the traffic load approaches or even gets higher than the system capacity, non of the three algorithms get good performance, since under excessive traffic load there is no policy that can guarantee the delay constraint.Note that there's a trade off between delay violation ratio and delay jitter according to the penalty coefficient $\mu$, a larger $\mu$ may cause the model to pay more attention on controlling the jitter rather than suppressing delay violation ratio.
 


Fig. \ref{fig:overall} shows the delay violation condition of different users using three algorithms under $G_u =$ 40, 52, 64 bits respectively, and the red line represents the delay violation ratio upper bound. When $G_u = 40$ bits, the traffic load remain under the system capacity upper bound, LGQP-IPS can perfectly achieve the delay constraint, while the two baseline algorithms exhibit user delay violations to varying extents. When $G_u = 52$ bits, the load is approaching the system capacity upper bound, all three algorithms show instances of user delay violations, but LGQP-IPS still maintains a
 relatively lower delay violation ratio. When $G_u = 64$ bits, the load is now over the system capacity upper bound, non of the three algorithms can guarantee the delay constraint, while LGQP-IPS still strives to minimize the average delay violation rate.


\section{Conclusion}
In this paper, a two-staged LGQP-IPS algorithm for delay-sensitive tasks in cellular downlink systems was proposed. In the first stage, a virtual queue was modeled to transform delay constraints into virtual queues to ensure compliance with delay requirements. In the second stage, a QMIX algorithm was introduced to intelligently schedule users and packets. 
Simulation results indicated that the proposed algorithm demonstrates lower delay jitter and higher compliance with delay constraints compared to baseline algorithms.

\bibliographystyle{IEEEtran}
\bibliography{IEEEabrv}

\begin{thebibliography}{10}
\providecommand{\url}[1]{#1}
\csname url@samestyle\endcsname
\providecommand{\newblock}{\relax}
\providecommand{\bibinfo}[2]{#2}
\providecommand{\BIBentrySTDinterwordspacing}{\spaceskip=0pt\relax}
\providecommand{\BIBentryALTinterwordstretchfactor}{4}
\providecommand{\BIBentryALTinterwordspacing}{\spaceskip=\fontdimen2\font plus
\BIBentryALTinterwordstretchfactor\fontdimen3\font minus \fontdimen4\font\relax}
\providecommand{\BIBforeignlanguage}[2]{{%
\expandafter\ifx\csname l@#1\endcsname\relax
\typeout{** WARNING: IEEEtran.bst: No hyphenation pattern has been}%
\typeout{** loaded for the language `#1'. Using the pattern for}%
\typeout{** the default language instead.}%
\else
\language=\csname l@#1\endcsname
\fi
#2}}
\providecommand{\BIBdecl}{\relax}
\BIBdecl

\bibitem{3gpp2021service}
{3rd Generation Partnership Project (3GPP)}, ``Service requirements for the 5g system (3gpp ts 22.261 version 16.14.0 release 16),'' 3GPP, 3GPP Standard TS 22.261, 4 2021.

\bibitem{yin2023joint}
Y.~Yin, B.~Liu, P.~Zhu, X.~Lyu, and Y.~Wang, ``Joint long-term energy efficient scheduling and beamforming design for urllc in cell-free mimo systems,'' \emph{IEEE Wireless Communications Letters}, 2023.

\bibitem{ewaisha2017optimal}
A.~E. Ewaisha and C.~Tepedelenlio{\u{g}}lu, ``Optimal power control and scheduling for real-time and non-real-time data,'' \emph{IEEE Transactions on Vehicular Technology}, vol.~67, no.~3, pp. 2727--2740, 2017.

\bibitem{neely2013dynamic}
M.~J. Neely and S.~Supittayapornpong, ``Dynamic markov decision policies for delay constrained wireless scheduling,'' \emph{IEEE Transactions on Automatic Control}, vol.~58, no.~8, pp. 1948--1961, 2013.

\bibitem{shafieirad2022meeting}
H.~Shafieirad and R.~S. Adve, ``On meeting a maximum delay constraint using reinforcement learning,'' \emph{IEEE Access}, vol.~10, pp. 97\,897--97\,911, 2022.

\bibitem{fan2021delay}
Q.~Fan, J.~Bai, H.~Zhang, Y.~Yi, and L.~Liu, ``Delay-aware resource allocation in fog-assisted iot networks through reinforcement learning,'' \emph{IEEE Internet of Things Journal}, vol.~9, no.~7, pp. 5189--5199, 2021.

\bibitem{elsayed2019reinforcement}
M.~Elsayed and M.~Erol-Kantarci, ``Reinforcement learning-based joint power and resource allocation for urllc in 5g,'' in \emph{2019 IEEE Global Communications Conference (GLOBECOM)}.\hskip 1em plus 0.5em minus 0.4em\relax IEEE, 2019, pp. 1--6.

\bibitem{gutierrez2017synchronization}
M.~Guti{\'e}rrez, W.~Steiner, R.~Dobrin, and S.~Punnekkat, ``Synchronization quality of ieee 802.1 as in large-scale industrial automation networks,'' in \emph{2017 IEEE Real-Time and Embedded Technology and Applications Symposium (RTAS)}.\hskip 1em plus 0.5em minus 0.4em\relax IEEE, 2017, pp. 273--282.

\bibitem{polyanskiy2010channel}
Y.~Polyanskiy, H.~V. Poor, and S.~Verd{\'u}, ``Channel coding rate in the finite blocklength regime,'' \emph{IEEE Transactions on Information Theory}, vol.~56, no.~5, pp. 2307--2359, 2010.

\bibitem{mao2017stochastic}
Y.~Mao, J.~Zhang, S.~Song, and K.~B. Letaief, ``Stochastic joint radio and computational resource management for multi-user mobile-edge computing systems,'' \emph{IEEE transactions on wireless communications}, vol.~16, no.~9, pp. 5994--6009, 2017.

\bibitem{rashid2020monotonic}
T.~Rashid, M.~Samvelyan, C.~S. De~Witt, G.~Farquhar, J.~Foerster, and S.~Whiteson, ``Monotonic value function factorisation for deep multi-agent reinforcement learning,'' \emph{Journal of Machine Learning Research}, vol.~21, no. 178, pp. 1--51, 2020.

\bibitem{zhu2022survey}
C.~Zhu, M.~Dastani, and S.~Wang, ``A survey of multi-agent reinforcement learning with communication,'' \emph{arXiv preprint arXiv:2203.08975}, 2022.

\bibitem{langer2020distributed}
M.~Langer, Z.~He, W.~Rahayu, and Y.~Xue, ``Distributed training of deep learning models: A taxonomic perspective,'' \emph{IEEE Transactions on Parallel and Distributed Systems}, vol.~31, no.~12, pp. 2802--2818, 2020.

\bibitem{samvelyan19smac}
M.~Samvelyan, T.~Rashid, C.~S. de~Witt, G.~Farquhar, N.~Nardelli, T.~G.~J. Rudner, C.-M. Hung, P.~H.~S. Torr, J.~Foerster, and S.~Whiteson, ``{The} {StarCraft} {Multi}-{Agent} {Challenge},'' \emph{CoRR}, vol. abs/1902.04043, 2019.

\end{thebibliography}


\end{document}